\documentclass[apj,numberedappendix]{emulateapj}
\usepackage{graphics,epsf}
\usepackage{amsmath}                
\usepackage{amsfonts}               
\usepackage{amssymb}                
\usepackage{epsfig}                 
\usepackage{graphicx}
\usepackage{float}
\usepackage{color}
\usepackage[para,online,flushleft]{threeparttable}

\newcommand{\km}{{~\rm km}}

\newcommand{\K}{{~\rm K}}
\newcommand{\erg}{{~\rm erg}}
\newcommand{\yr}{{~\rm yr}}

\newcommand{\nar}{{~\rm New Astronomy Reviews}}

\begin{document}

\title{Supernovae Ia in 2017: a long time delay from merger/accretion to explosion}
\author{Noam Soker\altaffilmark{1,2}}

\altaffiltext{1}{Department of Physics, Technion -- Israel Institute of Technology, Haifa 32000, Israel; soker@physics.technion.ac.il}
\altaffiltext{2}{Guangdong Technion Israel Institute of Technology, Shantou, Guangdong Province, China}

\begin{abstract}
I use recent observational and theoretical studies of type Ia supernovae (SNe Ia) to further constrain the viable SN Ia scenarios and to argue that there must be a substantial time delay between the end of the merger of the white dwarf (WD) with a companion or the end of mass accretion on to the WD and its terminal explosion.
This merger/accretion to explosion delay (MED) is required to allow the binary system to lead to a more or less spherical explosion and to prevent a pre-explosion ionizing radiation. Considering these recent results and the required MED, I conclude that the core degenerate scenario is somewhat more favorable over the other scenarios, followed by the double degenerate scenario. Although the single degenerate scenario is viable as well, it is less likely to account for common (normal) SN Ia. As all scenarios require substantial MED, the MED has turned from a disadvantage of the core degenerate scenario to a challenge that theory should overcome. I hope that the requirement for a MED will stimulate the discussion of the different SN Ia scenarios and the comparison of the scenarios to each other.
\end{abstract}


\section{INTRODUCTION}
\label{sec:intro}

Researchers over the years have considered six different scenarios to explain the evolution of carbon-oxygen white dwarfs (CO WDs) toward explosion as type Ia supernovae (SNe Ia; I will not consider scenarios that involve gravitational energy, like the quark nova scenario, QN-Ia, e.g., \citealt{Ouyedetal2015}).
\cite{TsebrenkoSoker2015a} and \citet{Soker2015} compare to each other the scenarios that involve binary stellar systems, describe the properties that distinguish one model from the other, and list their pro and cons. Before I turn in the coming sections to narrow down the viable scenarios, I describe  (by alphabetical order) each one of them in short.
\begin{enumerate}
\item
\textit{The core-degenerate (CD) scenario.} A CO WD merges with the core of a massive asymptotic giant branch (AGB) star during a common envelope evolution. The merger product has a mass of about the Chandrasekhar mass and explodes much later. This was developed as a separate SN Ia scenario in the last decade (e.g., \citealt{Kashi2011, Soker2011, Ilkov2012, Ilkov2013, Sokeretal2013, AznarSiguanetal2015}). In an earlier study \cite{Livio2003} considered the core-WD merger to be a rare event rather than the main SN Ia scenario. The scenario that was discussed by \cite{SparksStecher1974} and that involves the collision of a WD with the core does not belong to the CD scenario because most the the energy comes from the collapse to a neutron star and there is a remnant of a neutron star.
\item
\textit{The double degenerate (DD) scenario},
In this scenario the explosion occurs after two WDs merge (e.g., \citealt{Webbink1984, Iben1984}). An open question in this scenario is the time delay from merger to explosion. It is not clear in this scenario how long after merger explosion occurs (e.g., \citealt{LorenAguilar2009, vanKerkwijk2010, Pakmor2013, Levanonetal2015}).
\item
\textit{The 'double-detonation' (DDet) mechanism.}
According to this scenario a CO WD accumulates a helium-rich layer material on its surface. The detonation of the helium-rich layer ignites the CO WD (e.g., \citealt{Woosley1994, Livne1995, Shenetal2013}). I also include in this scenario cases where the donor star is a WD, so there are two degenerate stars in the system, e.g., the ``dynamically driven double degenerate double detonation'' channell that is studied by \cite{Shenetal2017}.
\item
\textit{The single degenerate (SD) scenario.} A non-degenerate stellar companion transfers hydrogen rich gas (or even helium rich gas, e.g., \citealt{Ruiteretal2009}) to the CO WD. Explosion takes place when the CO WD reaches close to the Chandrasekhar mass limit (e.g., \citealt{Whelan1973, Nomoto1982, Han2004}). Else, the CO WD explodes only after it loses some of its angular momentum after a long time (e.g., \citealt{Piersantietal2003, DiStefanoetal2011, Justham2011}), or after redistribution of angular momentum inside the WD even without losing angular momentum \citep{Justham2011}.
\item
\textit{The singly-evolved star (SES) scenario.}
Pycnonuclear reactions that require traces of hydrogen in a single CO WD drive a powerful detonation \citep{Chiosietal2015}.
This model requires a typical hydrogen mass fraction that is much larger than what is found in detailed calculations of CO WD evolution. For that I will not discuss this scenario anymore in the present study.
\item
\textit{The WD-WD collision (WWC) scenario.}
The two CO WDs collide at about their free fall velocity into each other, and by that ignite the nuclear reaction.
(e.g., \citealt{Raskinetal2009, Rosswogetal2009, LorenAguilar2010, Thompson2011, KatzDong2012, Kushniretal2013, AznarSiguanetal2013, AznarSiguanetal2014}).
\end{enumerate}

Each of the five binary scenarios has strong and weak points, as I summarize in Table \ref{tab:Table1} that is based on the work of \cite{TsebrenkoSoker2015a}. Beside the last three rows, the table is as presented by \cite{TsebrenkoSoker2015a}. The last three rows are my present estimate of the contribution of each scenario to the classes of common SNe Ia and of peculiar SNe Ia. The last three rows are the main subject of the present study. \cite{Sokeretal2014} compare the first four scenarios that are listed in Table \ref{tab:Table1} against some other observations (their Table 1).
\begin{table*}
\scriptsize
\begin{center}
  \caption{Confronting five SN Ia scenarios with observations}
    \begin{tabular}{| p{1.8cm} | p{2.5cm}| p{2.5cm}| p{2.5cm}| p{2.5cm} | p{2.5cm} |}
\hline  
{Scenario$^{[{\rm a}]}$}  & {Core Degenerate}    & {Double Degenerate} & {Double Detonation} & {Single Degenerate} & {WD-WD collision} \\

\hline  

 {Presence of 2 opposite Ears
in some SNR~Ia.}
  & Explained by the SN inside planetary nebulae (SNIP) mechanism.
  & Low mass Ears if jets during merger \citep{TsebrenkoSoker2013}.
  & No Ears are expected for He WD companion.
  & Ears by jets from accreting WD \citep{TsebrenkoSoker2013}.
  & No Ears are expected   \\

\hline  

 {$\approx 1M_\odot$ CSM in Keplers SNR }
  & The massive CSM shell might be a PN.
  & No CSM shell
  & Any CSM is of a much lower mass
  & Can be explained by heavy mass loss from an AGB donor. $^{[{\rm e}]}$
  &  No CSM shell   \\

\hline  
 {Main
 Scenario

 Predictions}
 & 1. Single WD explodes \newline2. Massive CSM in some cases (SNIP)
 & 1. Sufficient WD-WD close binaries \newline2. DTD~$\propto t^{-1}$
  & 1. Asymmetrical explosion \newline2. $M_{\rm WD} <1.2 M_\odot$
  & 1. Companion survives \newline2. $M_{\rm WD} \simeq M_{\rm Ch}$
  &  Asymmetrical \newline explosion   \\
\hline  
 {General

  Strong

 Characteristics}
 & {\textcolor[rgb]{0.00,0.59,0.00}{1. Explains some SN Ia with H-CSM \newline 2. Symmetric explosion}}
 & {\textcolor[rgb]{0.00,0.59,0.00}{Explains very well the delay time distribution (DTD)}}
 & {\textcolor[rgb]{0.00,0.59,0.00}{Ignition easily \newline achieved}}
 & {\textcolor[rgb]{0.00,0.59,0.00}{1. Accreting massive WDs exist \newline 2. Many explosions with $\sim M_{\rm Ch}$ }}
 & {\textcolor[rgb]{0.00,0.59,0.00}{1. Ignition easily achieved \newline 2. compact object}} \\
\hline  
{General

 Difficulties}
 & {\textcolor[rgb]{0.8,0.0,0.8}{More work on \newline 1. Ignition process\newline 2. DTD\newline 3. Merge during CE \newline 4. Find massive single WDs}}
 & {\textcolor[rgb]{0.8,0.0,0.8}{1. Ignition process \newline2. Inflated gas around merger product$^{[{\rm b}]}$ \newline 3. Asymmetrical explosion  }}
 & {\textcolor[rgb]{0.8,0.0,0.8}{Ejected He in some sub-scenarios}}
 & {\textcolor[rgb]{0.8,0.0,0.8}{1. Cannot account for DTD \newline2. CSM of PTF 11kx too massive}}
 & {\textcolor[rgb]{0.8,0.0,0.8}{Does not reproduce the DTD at \newline $t<0.4$~Gyr}}
                               \\
\hline  
{Severe

 Difficulties}
 & {\textcolor[rgb]{0.98,0.00,0.00}{}}
 & {\textcolor[rgb]{0.98,0.00,0.00}{}}
 & {\textcolor[rgb]{0.98,0.00,0.00}{1. $M_{\rm WD} < 1.2M_\odot$ (no Mn production) \newline2. Highly asymmetrical explosion}}
 & {\textcolor[rgb]{0.98,0.00,0.00}{1. Not enough systems \newline2. Companions not found \newline3. No hydrogen observed}}
 & {\textcolor[rgb]{0.98,0.00,0.00}{1. Can account for $<1\%$ of SN Ia \newline 2.Highly asymmetrical explosion  }}
\\
\hline  
{Fraction of SN~Ia }
 & $> 20 \% $
 & $ < 80 \% $
 & $< {\rm few} \times\% $ $^{[{\rm c}]}$
 & $0 \% $
 & $<1 \%$ $^{[{\rm d}]}$  \\
\hline  
\hline  
 \textbf{This study:} \newline MED$^{[{\rm f}]}$
 & {\textcolor[rgb]{0.00,0.59,0.00}{MED is built-in}}
 & Must include \newline MED$\ga 10^5 \yr$
 & {\textcolor[rgb]{0.98,0.00,0.00}{MED is problematic with helium accretion}}
 &  2+3 are solved for MED$\ga10^7\yr$; \newline {\textcolor[rgb]{0.98,0.00,0.00}{problem 1 stays}}
 & {\textcolor[rgb]{0.98,0.00,0.00}{MED is impossible}} \\
\hline  
 \textbf{This study:} \newline fraction of common SN Ia
 & $\ga 60 \%$
 & $ \la 30 \%; \quad$ mainly sub-Chandrasekhar SNe Ia
 & $0 \%$
 & $ \la 10 \%$
 & $0\%$ \\
\hline  
 \textbf{This study:} \newline fraction of peculiar SN Ia
 & $\approx  10\% $
 & ${\rm few} \times 10\% $
 & ${\rm few} \times 10\% $
 & ${\rm few} \times 10\% $
 &  $<1 \% $ \\
\hline  

     \end{tabular}
  \label{tab:Table1}\\
\end{center}
\begin{flushleft}
\small Notes:\\
The first eight rows are from \cite{TsebrenkoSoker2015a}. The last three rows are from this study. The contributions of the different scenarios to SNe Ia and peculiar SNe Ia populations are crude estimates.  \\
 \small \footnotemark[1]{Scenarios for SN Ia by alphabetical order; see text for details. } \\
 \small \footnotemark[2]{Disk originated material (DOM) around the merger product rules out explosion within several~$\times10 \yr $ of merger \citep{Levanonetal2015}.
 Such a late exploison will solve also the asymmetrical SNR problem.  } \\
 \small \footnotemark[3] {\cite{Piersantietal2013}; see also \cite{Papishetal2015}.} \\
 \small \footnotemark[4]{See Section 2 in first version of astro-ph (arXiv:1309.0368v1) of \cite{Sokeretal2014}}\\
 \small \footnotemark[5]{In some respects the formation of a shell with Ears in the SD scenario with an AGB star as the donor is similar to that in the CD scenario, as both are results of mass loss from an AGB star.} \\
 \small \footnotemark[6]{MED: merger/accretion to explosion delay time.}
\end{flushleft}
\end{table*}

It is not easy to distinguish between the different scenarios by the properties of the explosion itself (e.g., \citealt{Noebaueretal2017}).
The claim that early blue bump is a signature of an interaction with the companion, for example, was mentioned in recent years (e.g.,  \citealt{Caoetal2015, Marionetal2016} [but see \citealt{Shappeeetal2016} that find no excess light]; \citealt{Hosseinzadehetal2017}), but this can arise also in the DD scenario \citep{LevanonSoker2017}, and possibly by dense nickel clumps in the outer region in the CD scenario. Iron clumps (e.g., \citealt{TsebrenkoSoker2015b, Yamaguchietal2017}) and clumps in general (e.g., \citealt{Wilketal2018}) seem to be common in SNe Ia.
In any case, SNe Ia with early light excess are rare (e.g., \citealt{Biancoetal2011}).
Polarization at explosion is another properties that cannot (yet) be used to reveal the SN Ia scenario (e.g., \citealt{Mengetal2017}).

The origin of circumstellar matter (CSM) that leads to a SN Ia-CSM event, where the ejecta interacts with the CSM, is also controversial, i.e., through the SD scenario (e.g., \citealt{Sternbergetal2011, Maguireetal2013}) versus the CD scenario \citep{Soker2014NaDA}. Although many studies attribute the CSM to a mass loss from a giant star in the SD scenario, e.g., \cite{Bocheneketal2018} for a recent paper on X-ray emission from such an interaction, the CD scenario might do even better in accounting for the CSM, e.g., as in PTF~11kx \citep{Sokeretal2013}. \cite{Wangetal2014} however, claim that the SD scenario can account for substantial circumstellar hydrogen.
Relevant to this study is the $\approx 3 \times 10^4 \yr$ delay from a major CSM ejection and the explosion in Kepler SNR \citep{Katsudaetal2015}.

Even the delay time distribution (DTD; the time from star formation to explosion), does not always help to distinguish between the scenarios (e.g., \citealt{Heringeretal2017}), although in a recent study \cite{MaozGraur2017} claim that the DTD is as obtained in earlier studies, and hence the DD scenario fits well the observed DTD. In their review, \cite{Maozetal2014} prefer the DD scenario, but the debate on the SN Ia progenitors did not end with that review.

The newly used tool of very late times (beyond about 1000 day) observations of SNe Ia (e.g., \citealt{Seitenzahletal2009, Ropkeetal2012, Grauretal2018}) cannot yet distinguish between explosion models (e.g., \citealt{Grauretal2016}). This new tool seemingly has difficulties as different groups (e.g., \citealt{Dimitriadisetal2017, Shappeeetal2017}) arrive at different conclusions using this technique. \cite{Kerzendorfetal2017a} note that the unknown physics is likely the main part of this dispute.

The correlations between SNe Ia rates and galaxy properties, (e.g., \citealt{Grauretal2017a, Grauretal2017b}) might also be used to distinguish between different scenarios (e.g., \citealt{Ruiteretal2013}).
\cite{Shenetal2018} claim that this observation supports the sub-Chandrasekhar channel of the DD scenario. However, their qualitative fitting is not the final word yet. As well, it is possible that the dependence of SNe Ia properties on galaxy properties (e.g., \citealt{Childressetal2013}) might be related to the dependence of the fraction of the progenitor binary systems on metallicity (e.g., \citealt{Badenesetal2018}).

This seemingly stalemate situation has brought many researchers to the view that two or more scenarios might account for common SNe Ia (e.g., \citealt{WangHan2012, Hillebrandtetal2013, Rodneyetal2014, RuizLapuente2014, Childressetal2015, MaedaTerada2016}). In the present study I present the view that recent results allow some moves (so there is no stalemate) and by that can serve as tiebreakers.

The new items in the present study are the implementations of very new studies (2017 and on) to preclude the WWC scenario from the viable scenarios to common SNe Ia, and to argue that in the other four binary scenarios there must be a significant delay from the merger or accretion phase until the explosion itself.
The goal of the study is not to close the debate on the scenarios of SNe Ia, but rather to stimulate the discussion of the different scenarios and to encourage their comparison on an equal ground.

I build the paper around the requirement for a delay time from merger/accretion to explosion, going from the constraints on short merger/accretion to explosion delays (MEDs) to long MEDs. Along the different subsections of section \ref{sec:delay} I also list some other constraints on the different SN Ia scenarios. In section  \ref{sec:CDscenario} I discuss the CD scenario, the only scenario for which the MED is already accounted for.
I summarize my view in section \ref{sec:summary}, with the knowledge that my ranking of the likelihood of each scenario is controversial.

\section{THE MERGER-EXPLOSION DELAY (MED)}
\label{sec:delay}

There are several observational properties of (at least some) SNe Ia that require a time delay between the merger of the WD or the mass accretion process on to the WD to its terminal explosion. (This is not to be confused with the delay time from star formation to explosion, termed delay-time distribution, or DTD.)
I list the different constraints on the merger/accretion to explosion delay-time (MED) in the following subsections. The new parts are within the different subsections where I refer to papers from 2017 and on.

\subsection{Globally spherical type Ia supernova remnants}
\label{subsec:spherical}
\subsubsection{The severe problems of the WWC scenario}
\label{subsubsec:WWC}
The general spherical morphology of SN Ia remnants (SNRs Ia; e.g., \citealt{Lopezetal2009}), some who are known to be normal SN Ia, like the Kepler SN of 1604 (e.g., \citealt{RuizLapuente2017}), constrains the explosion to be more or less globally spherically symmetric. Bullets and jets are possible in some cases (e.g., \citealt{TsebrenkoSoker2015b}). This consideration almost precludes the WWC scenario as a viable scenario for normal SN Ia (e.g., \citealt{Mengetal2017}), as in the WWC scenario the explosion is highly non-spherical \citep{Kushniretal2013, PapishPerets2016}.

There is another severe problem for the WWC scenario in that its contribution to the total number of SNe Ia is very low (e.g., \citealt{Prodanetal2013}), about one per cent or less \citep{Hamersetal2013, Sokeretal2014, PapishPerets2016}. Two studies from 2017 did not help much the WWC scenario.
\cite{Toonenetal2017} conduct a thorough population synthesis study and conclude that the SN Ia rate from the WWC scenario might be of the order of $0.1 \%$ of all SNe Ia. \cite{Fangetal2017} find the merger rate of quadruple stellar systems to be a factor of $3.5 - 10$ higher than that in triple systems. But they start their study with already existing two WDs, not considering cases of merger of the two progenitors of the two WDs before the two WDs are formed. Based on the calculations in the above cited papers, my estimate is that their most optimistic case cannot bring the expected fraction of SNe Ia in the WWC scenario to more than one per cent. Even in these cases the explosion is more likely to be a peculiar SN Ia, e.g., because the SNR will be highly asymmetrical, unlike known SNRs Ia in the Galaxy.

Based on the above discussion, and despite the contrary conclusion in a very recent paper by \cite{Wygodaetal2018},  my view is that the WWC scenario does not contribute to common SNe Ia, and at best might contribute a very small fraction of peculiar SNe Ia (last two boxes of the last column in Table \ref{tab:Table1}).

\subsubsection{The short MED of the DD scenario}
\label{subsubsec:SDD}

The global spherical morphology of SNRs Ia puts a severe limit on the violent merger in the DD scenario because the merger process of the two WDs (e.g., \citealt{Kashyapetal2017}) and the explosion are highly non-spherical (e.g., \citealt{Pakmor2012, Tanikawaetal2015, vanRossumetal2016}). This implies that in the DD scenario the two WDs must first complete the merger process to leave behind a more or less spherical merger remnant when explosion eventually takes place.
 In the merger process one WD is destroyed and forms an accretion disk around the more massive WD. The merger is completed on the viscous time of the accretion disk which is  about tens of times of the dynamical time, that is several minutes.

The inferred nucleosynthesis of $^{55}{\rm Mn}$ in SNe Ia requires a high density at the center of the explosion, which in turn requires a massive exploding WDs, in at least $50\%$ of all SNe Ia (e.g., \citealt{Seitenzahletal2013a, Yamaguchietal2015, LeungNomoto2017}). The same holds for some other iron group elements (e.g., \citealt{Daveetal2017}) and for the $^{57}{\rm Ni}$ in SN~2011fe (\citealt{Dimitriadisetal2017, Shappeeetal2017}; but note that \citealt{Kerzendorfetal2017a} argue that they cannot infer the $^{57}{\rm Ni}$ mass with a precision that would allow the discrimination between models) and in SN~2012cg \citep{Grauretal2016}. To reach high density in the merger remnant the two merging WDs need to complete the merger process, namely, there must be a MED.

\cite{McWilliametal2017} study the most metal rich star in the Ursa Minor dwarf galaxy and show that the chemical composition is dominated by nucleosynthesis from a low-metallicity sub-Chandrasekhar mass SN Ia. Although it is quite possible that it was a sub-Chandrasekhar event, I would attribute the nucleosynthesis to a peculiar SN Ia rather than a normal SN Ia. \cite{CescuttiKobayashi2017} argue that sub-Chandrasekhar SNe Ia and/or peculiar SNe Ia, such as SNe Iax, are needed to explain the metalicity in Ursa Minor and similar dwarf galaxies.
Further support for sub-Chandrasekhar SNe Ia comes from the recent study by \cite{Shenetal2018}, who argue that the sub-Chandrasekhar channel of the DD scenario can account for the observation of fainter SNe Ia in old stellar populations.

Although the merger remnant can become spherical after several minutes, during the merger process the disk blows wind and/or jets that carry the extra angular momentum. As \cite{Levanonetal2015} point out, if the explosion occurs within about 1 day, this disc-originated matter (also termed DOM) will make the explosion looks like the exploding object has a size of $\ga 0.2 R_\odot$, contradicting observations of SN~2011fe \citep{Bloometal2012, PiroNakar2014}.

Over all, the spherical explosion of a compact object constraints the MED in the DD scenario to be longer than about one day. It should be noted already here that the claimed MED for the DD scenario does not change at all studies of the DTD because the DTD deals with much longer times scales $\gg 10^6 \yr$.

\subsubsection{The short MED of the DDet scenario}
\label{subsubsec:SDDet}

Recent models of the DDet scenario tend to take the He-donor companion to be a He WD (e.g., \citealt{Shenetal2017}). However, the He WD leads to asymmetrical explosion by two effects \citep{Papishetal2015}. In the first, that is applicable to all He WD companions, the He WD casts an ‘ejecta shadow’ behind it. At later times the outer parts of the shadowed side are fainter and the boundary of this side with the ambient gas is somewhat flat. This is not observed in SNRs. The second asymmetrical effect occurs for He WD donors of mass $\ga 0.4 M_\odot$. Such a donor must be closer to the exploding CO WD to transfer mass, and when the CO WD explodes it ignites the He WD. In this tripe-detonation scenario the outcome is a highly non-spherical explosion \citep{Papishetal2015}.
To explain the globally spherical SNRs Ia, the He donor in the DDet scenario should disappear. It is not clear how this can be achieved.
A non-degenerate helium star donor might lead to SN~Iax rather than to a normal SN Ia \citep{Wangetal2013}.

 Another strong constraint on the DDet scenario is the abundance of $^{55}{\rm Mn}$  and $^{57}{\rm Ni}$ as discussed in section \ref{subsubsec:SDD}.
The requirement to form $^{55}{\rm Mn}$ is a problem to the sub-Chandrasekhar DDet scenario that is not solved yet (e.g., \citealt{Shenetal2017}). One solution to this problem is a DDet scenario with a close to Chandrasekhar mass CO WD (e.g., \citealt{Jiangetal2017}).

\cite{Jiangetal2017} discuss SN~2016jhr (MUSSES1604D) and claim that the DDet explains this SN Ia. Their preferred model has a CO WD of mass of $1.38 M_\odot$ and an accreted helium mass of $0.03 M_\odot$. The accretion of this mass within a time $t_{\rm acc}$ results in a luminosity of  $L_{\rm WD} \simeq 1000 (t_{\rm acc}/10^5 \yr)^{-1} L_\odot$. As discussed in section \ref{subsec:ionization}, the accretion of this mass onto such a CO WD in less than about a million years will lead to over-ionization of the ambient gas in the case of Tycho's supernova, much as \cite{Woodsetal2017} argue for the SD scenario. In this model there is no room for a delay from accretion to explosion, as the helium supposes to be hot and ignites the CO WD.

I conclude with my view that the DDet scenario does not contribute to common SN Ia, but does contribute a small fraction of peculiar SN Ia (e.g., as \citealt{Wangetal2013} suggested for SN Iax), or to very energetic novae (if the WD survives).

\subsection{Circumstellar matter (CSM): The medium MED of the DD scenario}
\label{subsec:CSM}

There are several observations of early (within about 5 days after explosion) excess luminosity in  SN Ia light curves, both in normal SNe Ia (e.g., SN~2017cbv, \citealt{Hosseinzadehetal2017}) and in peculiar SNe~Ia (e.g., the SN~2002es-like event iPTF14atg, \citealt{Caoetal2015}).
Calculations explain this light excess by one of three mechanisms. A collision of the supernova ejecta with a companion in the SD scenario (e.g., \citealt{Kasen2010}), the collision of the ejecta with a previously ejected disc-originated matter (DOM) in the DD scenario \citep{LevanonSoker2017}, or an outer radioactive nickel clump \citep{Noebaueretal2017}, as expected in the CD scenario.

The point to make here is that this excess light seems to be rare. Namely, in most cases the different scenarios should avoid this light. This implies that in the SD scenario with a giant companion the explosion should take place after the giant has evolved to a small WD. \cite{Levanonetal2015} argue that in the DD scenario this implies that in most cases the MED should be longer than tens of years.

\subsection{Ionization of the Tycho's CSM}
\label{subsec:ionization}
\subsubsection{The long MED of the SD scenario}
\label{subsubsec:LSD}

Tycho's SNR is a well studied object, including studies of its CSM (e.g.  \citealt{Chiotellisetal2013, Zhouetal2016}). No companion that is expected in the SD scenario without MED has been found, and the new determined explosion site precludes the previous candidates \citep{XueSchaefer2015}. In general, the non detection of the expected bright companion to the exploding WD in SNRs according to the SD scenario (e.g., \citealt{SchaeferPagnotta2012, Pagnottaetal2014, Kerzendorfetal2017, SatoHughes2017, RuizLapuente2018}), as well as non-detection before explosion (e.g., \citealt{Kellyetal2014}) brought the idea of a very long delay from accretion to explosion, e.g., because rapid rotation keeps the structure overstable \citep{Piersantietal2003, DiStefanoetal2011, Justham2011, Boshkayevetal2014, Wangetal2014, Benvenutoetal2015}. This is termed the spin-up/spin-down model of the SD scenario. The spin-up/spin-down model helps the SD scenario in accounting for the very low CSM in most SNe Ia (e.g., \citealt{Horeshetal2012, Marguttietal2012, Marguttietal2014, Chomiuketal2016, DragulinHoeflich2016, Kunduetal2017}).
In a recent paper \cite{Kerzendorfetal2017} find no WDs in the SNR SN~1006 with an age of less than about $10^8 \yr$. They argue that this result is inconsistent with many spin up/down models of the SD scenario, as well as the process studied by \cite{ShenSchwab2017} where a surviving WD companion captures some $^{56}$Ni-rich SN
ejecta that energies the WD to blow wind and become luminous.
\cite{MengPodsiadlowski2013} argue that to account for SNe Ia interacting with the CSM the MED, in at least those SN Ia-CSM events, should be less than few$\times 10^7 \yr$

\cite{Johanssonetal2014} put a strong constraint on models of the SD scenario that include a long supersoft X-ray phase of the progenitor.
\cite{Grauretal2014} claim against the presence of a supersoft X-ray source in the $\approx 10^5 \yr$ prior to explosion of SN 2011fe, and by that limit the SD scenario to this SN Ia.
In a recent study \cite{Woodsetal2017} further constrain the duration of the MED. \cite{Woodsetal2017} find that the medium around the Tycho's SNR is not ionized, and from that rule out a steadily nuclear-burning white dwarf progenitor (that is a supersoft X-ray sources). They also rule out emission from an accretion disk around a Chandrasekhar-mass WD accreting at a rate of $\ge 10^{-8} M_\odot \yr^{-1}$ as is expected in a recurrent novae progenitor.

For a WD effective temperature of about $10^5 \K$, \cite{Woodsetal2017} constrain the luminosity of the WD to be $\la 10 L_\odot$ in the 100,000 years before explosion.
Their results limit the immediate progenitor of Tycho's supernova to be relatively faint.
\cite{Woodsetal2017} claim that the MED of the SD scenario must therefore be longer than about $10^5 \yr$. For a red giant companion the MED must be much longer to avoid detection in Galactic SNRs Ia.

\subsubsection{The long MED of the DD scenario}
\label{subsubsec:MDD}

We have seen that a short MED (less than about 100 years) is not possible for the DD scenario by the considerations of earlier sections. To reach a density for $^{55}{\rm Mn}$ formation the more massive WD, say of mass of $\approx 1 M_\odot$, should accrete more than $0.1 M_\odot$. This implies the release of $\ga 10^{49} \erg$ during the accretion process.
Even if only 10 per cent of this energy is released by a gas at a temperature of $\approx 10^5 \K$ the ionization will be more than allowed by the results of \cite{Woodsetal2017} for the Tycho's SNR.
 If the merger product blows a strong wind the photosphere is further out and the temperature is lower such that it causes no ionization,  but then the wind should have its own effect (M. van Kerkwijk, private communication).

\cite{TornambPiersanti2013} study the merger of a $0.8 M_\odot+0.7 M_\odot$ CO WDs in the frame of the DD scenario. They find that for about a million year the luminosity of the merger remnant is $\approx 10-100 L_\odot$ and its effective temperature is $\approx 10^5 \K$. As \cite{Woodsetal2017} show, this cannot be the progenitor of Tyco's SN, unless the merger remnant had a time to substantially cool.

Overall, I argue here that the MED of the DD scenario should extend in some cases (but not in all cases) up to at least several million years.

\section{THE CD SCENARIO}
\label{sec:CDscenario}

The DD, DDEt, SD and the WWC scenarios were not constructed originally with a significant MED. As we learned from section \ref{sec:delay}, observations impose long MED times on all binary scenarios. As the WWC can not include any MED, in the present study I exclude it from contributing to the population of common SNe Ia. The other scenarios must include substantial MED times, namely, orders of magnitudes longer than the dynamical time scale of a WD, and up to millions of years and more.

In the CD scenario, on the other hand, the MED is already accounted for \citep{Kashi2011, Ilkov2012, Ilkov2013}. For several years the MED was considered as a weak point of the CD scenario. In light of the understanding (that is summarized in section \ref{sec:delay}) that all scenarios require a long MED in some cases, this cannot be considered anymore as a disadvantage. The requirement for a substantial MED rather becomes now a challenge that theory should overcome.

Other recent studies seem to support the CD scenario (e.g., B. Schaefer, private communication; \citealt{Vink2016, RuizLapuente2018}). \cite{Cikotaetal2017} find that the polarization curves of some proto-planetary nebulae are similar to those observed along SNe Ia sight-lines. This, they argue, suggests that some SNe Ia explode during their post-AGB phase. The explosion inside a post-AGB wind, a planetary nebula, or a pre-planetary nebula, is termed SNIP \citep{TsebrenkoSoker2015a}, for a SN inside a planetary nebula \citep{DickelJones1985}.

But there are more challenges the CD scenario should overcome.
These (as summarized in Table \ref{tab:Table1}) include the merger process inside the common envelope and the eventual ignition of the WD. The same challenges are applicable to the long-MED DD scenario, but the merger is of two WDs due to gravitational waves rather than inside a common envelope.  For a recent study of igniting detonation see, e.g., \cite{GargChang2017}.

Another challenge is to find massive WDs before they explode. In the vicinity of the Sun all of them are single WDs (or have a companion at a wide separation) and most of them are expected to be very old and cold. There is a need to find what their expected radiation is, and to look for them.
On the positive side, it is possible that such WDs have already been found (e.g., \citealt{Ruedaetal2013, CoelhoMalheiro2014, Lobatoetal2016}).

Yet another challenge is the delay time distribution (DTD), i.e., the distribution of the rate of SNe Ia with time after star formation.
\cite{Wangetal2017} perform a detail population synthesis of the CD scenario, and conclude that it can account for no more than 20 per cent of all normal SNe Ia. They find that the main difference between their study and the earlier simpler study by \cite{Ilkov2013} is the amount of mass that the primary transfers to the secondary star in the first interaction between the two progenitors. \cite{Wangetal2017} find that the initially more massive star transfers less mass than assumed by \cite{Ilkov2013}.
Some other earlier studies also find the contribution of the CD scenario to be low, e.g., $<1\%$ \citep{MengYang2012} and $2-10 \%$ \citep{Zhouetal2015}.

My point is that more detail studies are required of both mass transfer and of the outcome of the common envelope evolution. The DD scenario requires the common envelope to bring two WDs very close together. I argue that in about an equal number of times or more, the massive WD and core (total mass of core and WD more than about $1.3 M_\odot$) will merge during the common envelope phase.

The CD scenario have several falsifiable predictions. The first and clearer one is the prediction that there is no close surviving stellar companion, as well as no companion just before explosion. The second one is that the exploding object has a radius of about $3000 \km$ or less. The third one is that in some cases there will be CSM of a mass of about $1 M_\odot$ \citep{Sokeretal2013}. As the WD-core merger is expected to take place for stars of an initial mass of $ \ga 4 M_\odot$, the massive CSM is not expected for SNe Ia in old stellar populations. The fourth prediction is that in young stellar populations where stars of initial mass of $\approx 4 M_\odot$ have finished their evolution but most SNe Ia are yet to take place, about one per cent of all WDs will have a mass close to the Chandrasekhar mass. The fifth prediction is that all WDs that explode through the CD scenario have a mass of about the Chandrasekhar mass. This mass is determined by a feedback mechanism during the merger process.

\section{SUMMARY}
\label{sec:summary}

In the present study I summarized arguments for a long delay time from the merger process in the CD, DD, and WWC scenarios, or from the completion of the accretion process in the DDet and the SD scenarios to the explosion itself (MED).
I concluded that all scenarios should have the possibility for a MED of more than about $10^5 \yr$. In a fraction of cases the delay might be shorter. The results are summarized in Table \ref{tab:Table1}.

As the WWC cannot have any delay from collision to explosion, it seems that this scenario does not contribute to common SN Ia. As well, the rate of collisions is much too low. It seems that the community should consider abandoning this scenario for the case of common SNe Ia. It might give rare cases of peculiar SNe Ia.

In the DDet scenario there are no considerations of any MED. In channels of the DDet scenario where the donor is a He WD even a long MED will not help because the WD donor will still be there at explosion and will cast a shadow in the SNR. This is not observed in SNRs.
The He WD might merge with the CO WD and not be there at explosion. But the rate of CO WD merging with a He WD is only about 15 per cent \citep{Liuetal2017} to 25 per cent \citep{Karakasetal2015} of the common SNe Ia rate. As well, this merger process might lead to a late electron capture transient event rather than a SN Ia \citep{Brooksetal2018}.
However, being similar to cataclysmic variables, many systems of CO WDs that accrete helium rich gas are expected to exist. These systems will contribute quite a lot to the population of peculiar SNe Ia (e.g., \citealt{Peretsetal2010, Waldmanetal2011, Perets2014, Crockeretal2017}).

Although the spin-up/spin-down model of the SD scenario that has a long MED can solve many of the problems of the traditional model of the SD scenario (where there is no MED), the majority of population synthesis studies show that the SD scenario can contribute about 10 per cent or less to the total population of SNe Ia (e.g., \citealt{Ruiteretal2009, Chenetal2014, Toonenetal2014, Liuetal2018a}; \citealt{Wangetal2017a} show that helium star donors do not help either). These events are more likely to be without MED, and contribute to the population of peculiar SNe Ia more than to normal SNe Ia. Even the new channel of the SD scenario that involves a common envelope as suggested by \cite{MengPodsiadlowski2017} does not increase the fraction by much.
This model of common envelope wind might form a massive CSM around a SN Ia in the SD scenario \citep{MengPodsiadlowski2017}.

The most significant claim of this study is that the DD scenario also requires a long MED in some of the cases.
By long I refer to a MED of at least $10^6 \yr$. As noted before, this MED does not change studies of the DTD that deals with delay times of $\gg 10^6 \yr$.
A long MED is not part of the traditional DD scenario, but was considered in different studies (e.g., \citealt{TornambPiersanti2013}).
The DD scenario seems to require that the total mass of the two WDs be close to the Chandrasekhar mass (e.g., \citealt{Danetal2015} and arguments earlier in the present study), as in the calculations of  \cite{TornambPiersanti2013}.
A problem of this scenario is that there are not enough binary WD systems with mass close to the Chandrasekhar mass \citep{BadenesMaoz2012}, although a recent study shows that this claim is not settled yet, and it might eventually turn out that there are enough massive WD-WD systems \citep{MaozHallakoun2017}. Still, even in these cases I claim that many systems should have a long MED.

Population synthesis studies have some difficulties in attributing all SNe Ia to the DD scenario when the total mass of the two WDs is close to the Chandrasekhar mass (e.g., \citealt{Ruiteretal2011, Toonenetal2012, Claeysetal2014}). Recent optimistic studies were conducted by, e.g., \cite{YungelsonKuranov2017} and \cite{Liuetal2018b}.
However, \cite{Liuetal2018b} assume that all stars are in binaries and a Galactic star formation rate of $5 M_\odot \yr^{-1}$, which is about 2.5 times the present star formation rate in the Galaxy (e.g., \citealt{ChomiukPovich2011}). For other galaxies they are short by a factor of 2-3 from the observed value.
\cite{YungelsonKuranov2017} also use a star formation rate that is 2.5 times larger than observed, as well as very efficient common envelope ejection.
The latest population synthesis results are a factor of $2-3$ lower than the observed SNe Ia rate in field galaxies (e.g., \citealt{Shenetal2018}).
Over all, even in these optimistic studies it seems that the DD scenario can account for no more than about half of all SNe Ia when more reasonable parameters are used. But this comes close to be within systematic errors, and population synthesis studies might in the near future explain all SNe Ia with the DD scenario (but still many should have a long MED).

The recent study of \cite{Shenetal2018} that supports sub-Chandrasekhar mass SNe Ia via the DD scenario (following earlier works, e.g., \citealt{vanKerkwijk2010}), namely, the sum of the two merging WDs is much below $1.4 M_\odot$, brings me to suggest that the main contribution of the DD scenario is to sub-Chandrasekhar mass SNe Ia. These cannot make the majority of the SNe Ia as I discussed earlier in the paper. This is the reason for my estimated fraction of SNe Ia via the DD scenario in Table \ref{tab:Table1}. As well, I expect the DD scenario to account for many peculiar SNe Ia (e.g., iPTF13asv, \citealt{Caoetal2016}).

 Despite the several challenges that studies should overcome to fully understand the CD scenario, it is my view that it is the most promising scenario for common SNe Ia (section \ref{sec:CDscenario}). It can also contribute to some peculiar SNe Ia. For example, in cases where either the core or the WD that merge during the common envelope phase are oxygen-neon (ONe) WD, the merger remnant will contain a rich ONe zone inside or outside. The thermonuclear outburst, if and when occurs, will be weaker than that of normal SNe Ia, e.g., as in SNe Iax.
I note that \cite{MengPodsiadlowski2014} argue that a CONe WD in a binary system with a main sequence star can form a SN Iax (2002cx-like SNe Ia).

Although it is not easy to ignite an ONe WD that accretes mass (e.g., \citealt{WuWang2017}), the situation might be different in the merger inside a common envelope of a CO core with an ONe WD, or an ONe core with a CO WD.

I end with repeating my call to consider all SN Ia scenarios in studies that aim to reveal the nature and origin of SNe Ia. With increasing number of new surveys and studies and those to come (e.g., \citealt{Chakrabortietal2016, GhoshWheeler2017, Isernetal2017, MulliganWheeler2017, RebassaMansergasetal2017}) it is important to consider all the SNe Ia scenarios, at least until the community comes to a consensus on the leading scenario(s).

I thank Or Graur, Wolfgang Kerzendorf, Pablo Loren-Aguilar, Xiangcun Meng, Ashley Ruiter, Pilar Ruiz-Lapuente, Marten van Kerkwijk, Bo Wang, and two anonymous referees for detail and helpful comments.
This research was supported by the Israel Science Foundation and by the E. and J. Bishop Research Fund at the Technion.


\label{lastpage}
\end{document}